\documentclass{IEEEtran}
\usepackage[utf8]{inputenc}
\usepackage[T1]{fontenc}
\usepackage{comment}
\usepackage{amsmath}
\usepackage{amsfonts}
\usepackage[english]{babel}
\usepackage{amsthm}
\usepackage{amssymb}
\usepackage{graphicx}
\usepackage{latexsym}
\usepackage{graphicx}
\usepackage{lipsum}
\usepackage{enumitem}
\usepackage{tikz, pgfplots}
\usepackage{eso-pic,xcolor}
\usepackage{overpic, scalerel}
\pgfplotsset{compat=1.18}
\IEEEoverridecommandlockouts
\newcommand{\bPhi}{\boldsymbol{\Phi}}

\usepackage{subcaption}
\usepackage{multicol}
\usepackage{hyperref}
\usepackage[hyphenbreaks]{breakurl}
\usepackage{cleveref}
\newcommand{\LL}{\scaleto{K}{6.0pt}}
\newcommand{\Lo}{\scaleto{1}{6.0pt}}

\newcommand{\ky}{\scaleto{Y^2_{k_l}}{6.5pt}}

\newcommand{\customref}[2]{\hyperref[#1]{#2}}

\newcommand{\lk}{\scaleto{{{k_l}}}{6.0pt}}

\newcommand{\mkm}{\scaleto{{k_l,1}}{6.5pt}}
\newcommand{\mk}{\scaleto{{k_l,m}}{6.5pt}}
\newcommand{\mkM}{\scaleto{{k_l,M}}{6.5pt}}

\newcommand{\mkl}{\scaleto{{k_{l'},m}}{6.5pt}}
\newcommand{\Ul}{\scaleto{k_{l'}}{5.5pt}}
\newcommand{\mkltot}{\scaleto{{k_{l'},m}}{6.5pt}}

\newcommand{\transpose}{\mkern-1mu\scaleto{\mathrm{T}}{3.5pt}}
\newcommand{\common}{\frac{\pi^2 L_{\frac{1}{2}}(-\kappa_h)^2L_{\frac{1}{2}}(-\kappa_g)^2}{16(\kappa_h + 1)(\kappa_g + 1)}}
\newcommand{\ml}{M}

\newcommand{\U}{\scaleto{k_l}{5.5pt}}
\newcommand{\UU}{\scaleto{k_l}{3.5pt}}
\newcommand{\UUU}{\scaleto{k_l}{4.5pt}}

\newcommand{\LFR}{\scaleto{KR}{5.0pt}}

\newcommand{\snr}{\mathrm{\gamma}}

\newcommand{\bla}[1]{\textcolor{black}{#1}}

\newcommand{\e}{\mathbb{E}}

\newcommand{\ris}{\scaleto{\mathrm{RIS}}{3.5pt}}
\newcommand{\risss}{\scaleto{\mathrm{RIS}}{3.0pt}}

\newcommand{\p}{\mathrm{P}}

\newcommand\scalemath[2]{\scalebox{#1}{\mbox{\ensuremath{\displaystyle #2}}}}

\usepackage[figurename=Fig.]{caption}
\newcommand{\sca}{0.9}
\newcommand{\xyfont}{\fontsize{9.0pt}{2pt}\selectfont}
\newcommand{\tickfont}{\fontsize{7pt}{2pt}\selectfont}

\newcommand{\majorgrid}{major grid style={gray!30}}
\newcommand{\minorgrid}{minor grid style={dotted,gray!50}}
\newcommand{\captionwidth}{\captionsetup{width=\linewidth}}
\usepackage{physics}

\title{Beyond Diagonal RIS-Aided Networks: Performance Analysis and Sectorization Tradeoff\vspace{2mm}}

\author{{Mostafa Samy,~\IEEEmembership{Student Member,~IEEE}, Hayder Al-Hraishawi,~\IEEEmembership{Senior Member,~IEEE},\\ Abuzar B. M. Adam,~\IEEEmembership{Member,~IEEE}, Konstantinos Ntontin,~\IEEEmembership{Member,~IEEE}\\ Symeon Chatzinotas,~\IEEEmembership{Fellow,~IEEE}, and Björn Otteresten,~\IEEEmembership{Fellow,~IEEE}\vspace{-7mm}}\\
\thanks{The authors are with the Interdisciplinary Centre for Security, Reliability and Trust (SnT), University of Luxembourg, Luxembourg. Corresponding author: \emph{Mostafa Samy (mostafa.samy@uni.lu)}.}
\thanks{This work was supported by the Luxembourg  National Research Fund (FNR) through the AFR Project CEP-MBD-CRIS, Grant reference 17974844.}
\thanks{This work in part was submitted to the the IEEE 25th International Workshop on Signal Processing Advances in Wireless Communications (SPAWC), 2024 \cite{Mostafa2024WS}}

}

\begin{document}
\pagenumbering{arabic}
\maketitle
\begin{abstract} 
Reconfigurable intelligent surfaces (RISs) have emerged as a spectrum- and energy-efficient technology to enhance the coverage of wireless communications within the upcoming 6G networks. Recently, novel extensions of this technology, referred to as multi-sector beyond diagonal RIS (BD-RIS), have been proposed, where the configurable elements are divided into $L$ sectors $(L \geq 2)$ and arranged as a polygon prism, with each sector covering $1/L$ space. 
This paper presents a performance analysis of a multi-user communication system assisted by a multi-sector BD-RIS operating in time-switching (TS) mode. Specifically, we derive closed-form expressions for the moment-generating function (MGF), probability density function (PDF), and cumulative density function (CDF) of the signal-to-noise ratio (SNR) per user. Furthermore, exact closed-form expressions for the outage probability, achievable spectral and energy efficiency, symbol error probability, and diversity order for the proposed system model are derived. To evaluate the performance of multi-sector BD-RISs, we compare them with the simultaneously transmitting and reflecting (STAR)-RISs, which can be viewed as a special case of multi-sector BD-RIS with only two sectors. Interestingly, our analysis reveals that, for a fixed number of configurable elements, increasing the number of sectors improves outage performance while reducing the diversity order compared to the STAR-RIS configuration. This trade-off is influenced by the Rician factors of the cascaded channel and the number of configurable elements per sector. However, this superiority in slope is observed at outage probability values below $10^{-5}$, which remains below practical operating ranges of communication systems. 
Additionally, simulation results are provided to validate the accuracy of our theoretical analyses. These results indicate that increasing the number of sectors in multi-sector BD-RIS-assisted systems significantly enhances performance, particularly in spectral and energy efficiency. For instance, our numerical results show a notable $182\%$ increase in spectral efficiency and a $238\%$ increase in energy efficiency when transitioning from a 2-sector to a 6-sector configuration.
\end{abstract}

\begin{IEEEkeywords}
Beyond diagonal reconfigurable intelligent surface (BD-RIS), full-space coverage, multi-sector RIS, performance analysis, time switching mode.
\end{IEEEkeywords}

\section{Introduction}
The concept of the smart radio environment has been a focal point in recent discussions on the evolution of 6G wireless communications and the shaping of future wireless networks. Central to this visionary landscape is the transformative potential of reconfigurable intelligent surfaces (RISs) \cite{Wu2020}. These innovative surfaces provide a paradigm shift in configuring the propagation conditions for information-bearing signals, leading to new challenges in the design of wireless communications systems. Specifically, these surfaces, consisting of arrays of closely spaced elements, are designed to manipulate electromagnetic waves by controlling the individual phase, amplitude, and polarization of the elements \cite{Basar2019}. By strategically adjusting these properties, RISs can effectively steer signals, mitigate interference, extend coverage, and enhance both spectral and energy efficiency within wireless networks \cite{Kisseleff2020}. Their programmable nature promises to ameliorate wireless communication by offering high adaptability and flexibility in optimizing signal propagation and system performance \cite{Renzo2020}.

Several types of RIS architectures have emerged to cater to the evolving needs of communication systems. 
For instance, stacked intelligent metasurfaces (SIM) represent a notable advancement, wherein multiple layers of reconfigurable elements are stacked to achieve enhanced beamforming capabilities and more intricate wavefront manipulation \cite{Jiancheng2023}. Further, simultaneously transmitting and reflecting (STAR) RISs introduce a new concept by supporting both reflective and transmissive modes concurrently. This innovation extends the scope of RIS functionality, enabling full-space coverage and dynamic adaptation to varying channel conditions \cite{liu2021star}. 
Beyond-Diagonal RIS (BD-RIS) technology further pushes the boundaries of RIS capabilities by relaxing the constraints of traditional surfaces with diagonal phase shift matrices \cite{bruno2}. BD-RISs enable more flexible beam management and improved spatial diversity through interconnection among surface elements beyond the conventional RIS structure. These advancements emphasize the continuous evolution of RIS technology towards more sophisticated, versatile, and high-performance communication solutions.

From an implementation standpoint, RISs can be categorized  into three architectures based on circuit topology: single-, group-, and fully-connected architectures \cite{scattering}. The single-connected design, typical of traditional RISs, features non-interconnected components and has been extensively studied in existing literature. In contrast, group- or fully-connected designs, where subsets or all RIS elements are interconnected, offer enhanced potential for exploiting channel diversity. These interconnected designs hold promise for substantial improvements in system spectral and energy efficiency gains \cite{bruno2}. These advanced architectures, termed as BD-RISs, allow the phase shift matrix representing the reflective coefficients of the RIS to extend beyond a diagonal structure. Nevertheless, the adoption of group/fully-connected architectures in BD-RISs introduces increased circuit complexity and control overhead. 
To address this challenge, the concept of \emph{multi-sector BD-RIS} was introduced in \cite{bruno3}, leveraging group-connected reconfigurable impedance networks and antenna array arrangements, as illustrated in Fig. \ref{sec:system_model}. The multi-sector BD-RIS offers flexible deployment options and is particularly advantageous in millimeter wave scenarios and cell-free networks. In this context, STAR-RIS can be viewed  as a particular case of a group-connected architecture, specifically  when the group size is set to $2$.

\subsection{BD-RIS Operating Protocols}
For versatile operation of multi-sector BD-RIS, two key protocols can be employed: energy splitting (ES) and time switching (TS).
\begin{itemize}
    \item \textbf{Energy Splitting (ES)}: This protocol divides the incoming signal into segments corresponding to the number of sectors in the BD-RIS. This allows for simultaneous processing across multiple elements, simultaneously serving different users within each sector. However, ES comes with increased complexity  as it requires additional circuitry for rectification and power management, leading to more challenging operation and control. Additionally, beamforming optimization with ES necessitates joint optimization of coefficients across all elements, significantly increasing the algorithmic complexity.

    \item \textbf{Time Switching (TS)}: In contrast, the TS protocol operates in the time domain, sequentially serving different users across BD-RIS sectors in designated time intervals. This temporal approach ensures efficient resource allocation and utilization. A significant advantage of TS is its simplicity as it features uncoupled beamforming coefficients, which simplifies optimization processes and reduces computational overhead. Furthermore, TS facilitates interference-free communication by serving only one user per time slot in each sector. This effectively mitigates inter-user interference while enhancing communication reliability, thereby yielding improved power reduction, as demonstrated in \cite{timeswitch}. Consequently, TS is well-suited for unicast applications, preventing degradation of communication rates due to inter-user interference \cite{timeswitch}.

\end{itemize}
Given the advantages of the TS protocol, particularly its simplicity and interference mitigation capabilities, we adopt it for operating the multi-sector BD-RIS-assisted network in this paper.

\subsection{Related Works}
Extensive research efforts have been dedicated to exploring the capabilities and potential applications of RISs across various domains \cite{Samy2024}. In recent literature, studies have investigated the design, optimization, and deployment strategies of different types of BD-RIS architectures, reflecting the growing interest in this innovative technology. For wideband communication systems aided by BD-RIS, \cite{wideband_2} introduced a model to account for the response fluctuations of BD-RIS across signals of varying frequencies. Additionally, \cite{wideband_1} optimized the BD-RIS configuration to maximize channel capacity, assuming linear variations in BD-RIS phase-shifts across subcarriers. Reference \cite{A_Low_Complexity_BD} proposed a two-stage beamforming approach to jointly optimize the active beamforming at the base station (BS) and passive beamforming at the BD-RIS in order to enhance the sum-rate for a multi-user BD-RIS-assisted network. In \cite{Multi_band}, the authors presented a generalized frequency-dependent reflection model for configuring fully-connected and group-connected RISs within multi-band multi-cell multiple-input multiple-output (MIMO) networks.

Furthermore, the BD-RIS scattering matrix for an integrated sensing and communication (ISAC) system was studied in \cite{ISAC_BD_RIS} to maximize the weighted sum of the signal-to-noise ratio (SNR) at both the radar receiver and communication users. Reference \cite{Closed_Form} optimized the BD-RIS scattering matrix to maximize the received signal power for single-user single-input single-output (SISO) systems, single-user MIMO and multi-user multiple-input single-output (MISO) aided BD-RIS systems. The work in \cite{Discrete_BD_RIS} optimized the group- and fully-connected BD-RIS scattering matrix with discrete values, where the RIS impedance matrix is independently discretized. The work in \cite{cellfree}  investigated a scenario where a cell-free massive MIMO system aided by BD-RIS serves two groups of users; information users and energy users for simultaneous wireless information and power transfer (SWIPT) systems. 
In \cite{Maximizing_SE_EE}, the performance of various RIS technologies, including regular (reflective and passive), STAR, and multi-sector BD-RIS, was investigated in multi-user MIMO OFDM broadcast channels.

Moreover, \cite{localized} proposed the concept of distributed BD-RIS, where the elements are spread over a wide region, contrasting with localized RIS. Reference  \cite{BD_RIS_UAV} developed a BD-RIS-UAV system, deploying multiple BD-RISs on unmanned aerial vehicles (UAVs) within Mobile Edge Computing (MEC) networks. \cite{dualfunction} introduced a dual-function radar-communication (DFRC) aided BD-RIS system to enhance communication capacity and sensing precision, thereby improving coverage for both functions.
For the grouping strategy of the BD-RIS elements in the group-connected architecture, a static and dynamic method  based on channel statistics and the channel state information (CSI) adaptation were proposed in \cite{Static_group} and \cite{dynamic_group}, respectively. Similarly, a grouping scheme for BD-RIS elements is developed in \cite{Group_RSMA}, considering user scheduling and the optimal rate-splitting power-splitting ratio for BD-RIS aided rate-splitting multiple access (RSMA) within multi-user MIMO systems.

\subsection{Motivation and Contributions}

From the aforementioned studies, it is evident that important progress has been made in optimizing BD-RIS  communication networks. However, there is still a gap in offering a thorough performance analysis for multi-sector BD-RIS across different fading scenarios  and incorporating multi-user communications. This analysis is essential for establishing a  wide-ranging framework for multi-sector BD-RIS deployment, including spectrum utilization and energy efficiency. This paper aims to explore these aspects by investigating the potential of increasing the number of sectors within the BD-RIS architecture while maintaining a fixed number of reconfigurable elements. We seek to provide valuable insights into the spectral and energy efficiency of these systems. Through in-depth analyses, we derive closed-form expressions to quantify system performance metrics, revealing the intricate interplay between sectorization, reconfigurable element density, and overall efficiency. 
Ultimately, this work endeavors to provide tools to assist in wireless network design, thereby improving spectral and energy efficiency.

This paper studies the performance of a  multi-sector BD-RIS-aided network to explore their potential performance enhancements. To achieve this, we adopt a cell-wise single-connected multi-sector BD-RIS architecture, similar to \cite{bruno3}, to ensure analytically tractable closed-form expressions. This architecture is based on a group-connected reconfigurable impedance network with a group size equal to $L$  \cite{bruno3}.
This analysis of the multi-sector BD-RIS across different fading channels has not been previously studied in the open literature. Specifically, we investigate a multi-sector BD-RIS-assisted network comprising $L$ sectors, each of which containing $M$ configurable elements. Our study provides exact closed-form expressions for the optimal SNR statistics, including probability density function (PDF), cumulative distribution function (CDF), and moment generating function (MGF). 
The key contributions of this paper can be outlined as follows.
\begin{enumerate}
\item  Investigate a multi-sector BD-RIS-aided  communication system operating with the TS  protocol.  Within this framework, we explore diverse performance metrics while keeping the number of configurable elements at the BD-RIS constant. 
We analyze the impact of increasing the number of sectors, consequently reducing the number of elements per sector, in comparison to configurations with fewer sectors but more elements per sector. We also include the STAR-RIS architecture  as a reference point, treating it as a special instance of the multi-sector BD-RIS when $L=2$. 

\item Derive exact closed-form expressions of the outage probability per user over the Rician fading channels. Leveraging the gamma matching method, we provide  expressions for various important statistics, including the CDF, PDF, and MGF. Additionally, we compute the asymptotic outage probability for high-SNR and determine the associated diversity orders. 

\item Using an MGF-based approach, we present a closed form expression for the symbol error probability (SEP) of the studied system for binary phase shift keying (BPSK) modulation scheme serving as a representative example. We also perform a high-SNR asymptotic analysis to gain further insights into SEP performance.

\item To analyze the trade-off between the number of sectors in  BD-RIS and the achievable spectral and energy efficiency gains under the TS protocol, we derive end-to-end spectral and energy efficiency expressions for this system model. Our analysis incorporates the hardware power consumption at all system components, including the transmitter, receiver, and BD-RIS elements.

\item Provide comprehensive numerical simulations to substantiate our analytical findings and obtain practical insights into system design considerations.
\end{enumerate}

\begin{figure*}
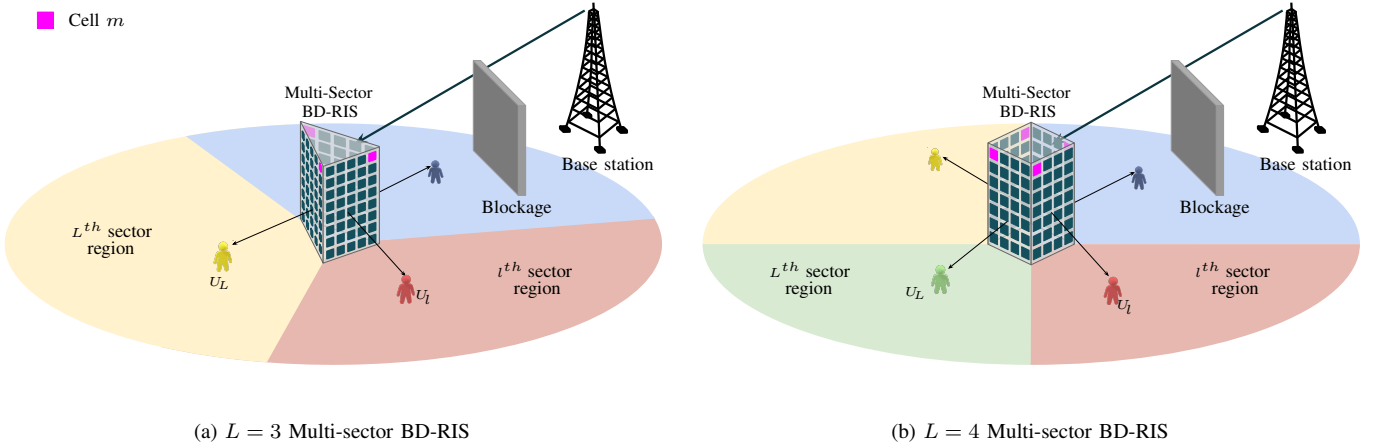

\begin{multicols}{2}
   \input{System_model/model1}\par 
   \input{System_model/model2}\par
    \end{multicols}
    \renewcommand{\figurename}{\footnotesize Fig}
      \caption{\footnotesize Multi-sector BD-RIS-aided multi-user communication system}
      \label{fig1}
\end{figure*}

\subsection{Paper organization}
The rest of the paper is organized as follows: The considered system model is described in Section \ref{sec:system_model}. In Section \ref{sec:Performance Analysis} performance analysis is carried out to obtain closed form-expressions for the PDF, CDF, outage probability, achievable rate, and diversity order. In Section \ref{sec:numerical}, numerical results are provided. Finally, conclusions are drawn in Section \ref{sec:conc}.

\subsection{Notation}
$A^T$  denotes the transpose of $A$. The notation $ A \sim \mathcal {CN} (\mu_A,\sigma^2_A)$ denotes that $A$ is a circularly symmetric complex Gaussian distributed with $\mu_A$ mean and $\sigma^2_A$ variance. Further, diag$(\boldsymbol{A})$
describes a vector with elements equal to the diagonal elements
of $\boldsymbol{A}$. The notation $\mathrm{Pr}[.]$ is the probability operation. The PDF of a random variable (RV) $X$ is denoted by $f_X(\cdot)$, and its CDF is denoted by $F_X(\cdot)$, respectively.
$\mathbb E[.]$ and $\mathbb{V}\mathrm{ar}[.]$ are the expectation
and variance operators, respectively.

\section{System Model}\label{sec:system_model}
As shown in Fig. \ref{fig1}, we consider a multi-sector BD-RIS-assisted communication system that consists of a BS and a multi-sector BD-RIS with $L$ sectors to serve $K$ users distributed randomly across the coverage areas of these sectors. We denote $\mathcal{K} = \{1,\dots,K\}$, and $\mathcal{L} = \{1,\dots,L\}$ as the set of indices of users, and sectors, respectively. This system operates under the assumption that direct links between the BS and users are obstructed due to severe blockage or the unfavorable
propagation environment. Moreover, it is assumed that the BS as well as $U_{\U}$ are equipped with a single antenna\footnote{This study focuses on analyzing the performance of multi-sector BD-RIS-assisted communications. Future work will extend this analysis to scenarios involving multiple antennas to further enhance system performance.}.
We adopt the TS protocol for the multi-sector BD-RIS configuration, where each user $U_{\U}$ is accommodated in a dedicated orthogonal time slot, with its corresponding sector activated concurrently.
To comprehend the concepts and terminology pertaining to multi-sector BD-RIS, we provide next a concise overview of its fundamental operational principles.
%

\subsection{Multi-Sector BD-RIS: Model and Setup}
The multi-sector BD-RIS is modeled as a polygon that consists of $L$ identical sectors and an $M$-cell RIS  \cite{bruno3}. We denote $\mathcal{M} = \{1,\dots,M\}$, and $\mathcal{K}_l=\{1,\dots,K_l\}$ as the set of indices of cells, and users located in the coverage region of sector $l,\forall l \in \mathcal{L}$ with $0 < K_l < K$, and $\sum_{l}K_{l}=K$, respectively. Based on the main characteristics of multi-sector BD-RIS,
the $M$-cell RIS is modeled as $\ml$ elements connected to an
$ML$-port reconfigurable impedance network, and uniformly divided into the $L$ sectors. 
Each cell consists of $L$ elements deployed at the side of the polygon while covering the whole azimuth space. In particular, cell $m$ includes elements in the set $\mathcal{L}_m=\{m, M+m, \ldots, (L-1)M+m\}$ \cite{bruno3}. The elements are interconnected by reconfigurable impedance components, and thus, enable the support of the multi-sector mode. Specifically, sector $l$ covers $1/L$ space containing elements which belongs to $\mathcal{M}_l = \{(l-1)M+1,...,lM\}$, $\forall l \in \mathcal{L} = \{1, \ldots, L\}$. 
The BS is located within the coverage
of sector $1$ of the multi-sector BD-RIS.
Moreover, the BS is out of the coverage of the
uni-directional radiation pattern of sector $l$ of the multi-sector BD-RIS, $\forall l \in L, l\neq 1,$ \cite{bruno3}. Therefore, the multi-sector BD-RIS involves $L$ matrices denoted as $\bPhi_{l} \in\mathbb{C}^{M \times M}$ for each sector $l$, where the constraint of
matrices $\bPhi_{l}$ is represented by the following equation \cite{bruno3} 
\begin{align}\label{totphase}
\sum_{l} \bPhi_{l}^{^{H}}\bPhi_{l}=\boldsymbol{I}_{M}.
\end{align}

Note that, based on the cell-wise architecture, $\bPhi_{l}$ in (\ref{totphase}), \(\forall l \in \mathcal{L}\) can be modeled as diagonal matrices as follows \cite{bruno3}
\begin{align}\label{main}
\bPhi_{l} = \text{diag}(\phi_{(l-1)M+1}, \ldots, \phi_{lM}), \quad \forall l \in \mathcal{L}, 
\end{align}
In the case of cell-wise single-connected architecture, constraint (\ref{main}) can be expressed as \cite{bruno3}
\begin{align}\label{Maximization}
\sum_{i\in m} |\phi_i|^2 = 1,~\forall m \in \mathcal{M}.
\end{align}
Assuming phase shifters with infinite resolution, where both the amplitude and phase shift can take continuous values, (\ref{Maximization}) can be reformulated as:
\begin{equation} \label{fourconst}
\begin{aligned}
&~~~\quad  \phi_i = \sqrt{\beta_i e^{j\theta_i}},
\forall i \in \mathcal{L}_m, \forall m \in \mathcal{M},\\
&~~~\quad  0 \leq \beta_i \leq 1, 
\forall i \in \mathcal{L}_m, \forall m \in \mathcal{M},  \\
&~~\quad  \sum_{i\in m} \beta_i = 1, 
\forall m \in \mathcal{M}, \\
&~~~\quad  0 \leq \theta_i < 2\pi,  
\forall i \in\mathcal{L}_m,  \forall m \in \mathcal{M}.
\end{aligned}
\end{equation}

\subsection{Channel Model}
The channel vectors between the BS and the sector of the BD-RIS facing the BS, and the BD-RIS and $U_{k_l}$ are denoted by $\mathbf{h} \in \mathbb{C}^{M\times 1}$ and $\mathbf{g}_{k_l} \in \mathbb{C}^{1\times M}$, respectively, where $\mathbf{h}=[h_1, ..., h_m, ..., h_M]^{\mathrm{T}}$, while the channel vector between the $l^{th}$ sector of the BD-RIS and $U_{\U}$ is denoted as $\mathbf{g}_{k_l} \in \mathbb{C}^{1\times M}$, where and $\mathbf{g}_{k_l}=[g_{\mkm}, ..., g_{\mk}, ..., g_{\mkM}]$.
In this, ${h}_{m},$ and $g_{\mk}$ capture the corresponding small-scale fading coefficients. Further, Rician fading is assumed for all channels, and thus, the small-scale coefficients to and from the BD-RIS $m^{th}$ cell are $h_m = \zeta_m e^{-j \vartheta_m}$ and $g_{\mk} =\xi_{\mk} e^{-j \upsilon_{\mk}}$, respectively, where $\zeta_m$ and $\xi_{\mk}$ are the amplitude coefficients, while $\vartheta_m$ and $\upsilon_{\mk}$ are the phase shifts.
Further, we denote $\alpha_{\U}$ as the cascaded channel large-scale fading coefficient expressed as \cite{bruno3}:
\begin{align}\label{large_scale_fading}
\alpha_{\U}=\frac{\lambda^{4}G_{t}G_{r}}{4^{3}\pi^{4}d_{\ris}^{\eta_{\risss}}d_{\U}^{\eta_{\UUU}}\qty(1-cos\frac{\pi}{L})^{2}},
\end{align}
where $d_{\ris}$ and $d_{\U}$ are the distances between the BS and the BD-RIS, and the BD-RIS and $U_{k_l}$, respectively, while $\eta_{\ris}$ and $\eta_{\U}$ are the corresponding path-loss exponents. In (\ref{large_scale_fading}) $\lambda$ represents the wavelength of transmit signal. 
%
%
Moreover, since we are focusing on studying the theoretical limits of this system model, and considering the channel estimation methods that have been recently developed in \cite{Static_group} and \cite{dynamic_group}, it is assumed that the BS can acquire accurate CSI. 
\subsection{Signal Model}
The received signal at user $U_{\U}$ randomly located within the $l^{th}$ sector coverage region of the BD-RIS is given by
\begin{align}
y_{\U} =\mathbf{g}_{\U}\boldsymbol{\Phi}\mathbf{h}\sqrt{p{\U}\mkern2mu\alpha_{\U}}s_{\U} + n_{\U},
\end{align}
where $s_{\U}$ is the message of $U_{\U}$, satisfying $\mathbb E\left[|s_{\U}|^2\right]=1$ \cite{Hayder2016}, 
$p_{\U}$ is the transmit power, 
and $n_{\U}\sim \mathcal {CN} (0,\sigma^2_{n_{\UU}})$ is the additive white Gaussian noise (AWGN) at $U_{\U}$.
Therefore, the received SNR of $U_{\U}$ is given by
\begin{equation}\label{eq:user_SNR}
\gamma_{\U}= \rho_{\U}|\mathbf{g}_{\U}\boldsymbol{\Phi}\mathbf{h}|^2\mkern2mu\alpha_{\U},
\end{equation}
where $\rho_{\U}=\frac{p_{k}}{\sigma_{n_{\UU}}^2}$ is the transmit SNR per user. Accordingly, the system overall spectral efficiency can be given as
\begin{align}\label{rate}
\mathcal{R}=\frac{\Lo}{\LL} \sum_{l=1}^{L} \sum_{k=1}^{K_{l}}\log_2 \qty(1 + \gamma_{\U}),
\end{align}
where the pre-log factor $\frac{1}{K}$ is included to account for the division of the total available time among the $K$ users, as dictated by the TS protocol.
\section{Performance Analysis}\label{sec:Performance Analysis}
In this section, our investigation begins with identifying the optimal phase shift configuration for the multi-sector BD-RIS to maximize its cascaded channel gains. Subsequently, we conduct statistical analyses for the optimal received SNR, deriving closed-form expressions for the PDF, CDF, and MGF of the multi-sector BD-RIS-aided system.
Moreover, we study various system performance metrics, including outage probability, achievable rate, diversity order, symbol error probability, and its asymptotic analysis. To comprehensively explore these aspects, we derive closed-form expressions, providing a thorough examination of the system behavior.  
\subsection{Optimal BD-RIS Configuration}
Since TS is utilized, where each user is served exactly once in a time, the maximum received signal power for $U_{k_l}$ within the coverage of sector $l'$, $\forall l' \in L$ at the time period $1/K$, can be achieved by activating its corresponding sector \cite{bruno3}. In this case, we have
\begin{align}\label{brunoadjus}
\mathbf{\Phi}_l =
\begin{cases}
\text{diag}(\phi_{(l-1)M+1}, \ldots, \phi_{lM}), &  l = l', \\
0, &  l \neq l',
\end{cases}
\end{align}
with $|\phi_j| = 1$, $\forall j \in M_{l'}$. Hence. the optimal phase-shifts design for $\boldsymbol{\phi}_{l'}$ can be given as follows \cite{bruno3}
\begin{align}\label{optimums}
\phi_{(l'-1)M+m} = -\angle [\mathbf{h}^{\transpose}]_m[\mathbf{g}_{\Ul}]_m
\end{align}
Upon implementing the optimal phase shifts $\phi_{(l'-1)M+m}$ from (\ref{optimums}), the cascaded channel for $U_{\Ul}$ can be expressed as:
\begin{align}
\big|\mathbf{h}^{\transpose}_{\bar{l}}
\mathbf{\Phi}^{_{~}}_{l}
\mathbf{g}_{\Ul}\big|^2=
& \bigg|\sum_{m=1}^{M} h^{_{~}}_m g^{_{~}}_{\mkl}\bigg|^2\nonumber\\
=& \bigg(\sum_{m=1}^{M}|\zeta^{_{~}}_m||\xi^{_{~}}_{\mkl}|\bigg)^2\nonumber\\
=& \bigg(\sum_{m=1}^{M}Y_{\mkltot}\bigg)^2=Y_{\Ul}^2.
\end{align}
\subsection{Statistics of the Optimal Received SNR}
We utilize the moment matching method to derive closed-form expressions characterizing the distribution of $\snr_{\U}$. This method is widely employed to approximate complex distributions \cite{Atapattu2020,gammdist2,gammadist3,mostafa2023} by matching the moments of the target distribution with those of a simpler distribution. Based on this method, the following Lemma gives the CDF, PDF and MGF of $\snr_{\U}$. 

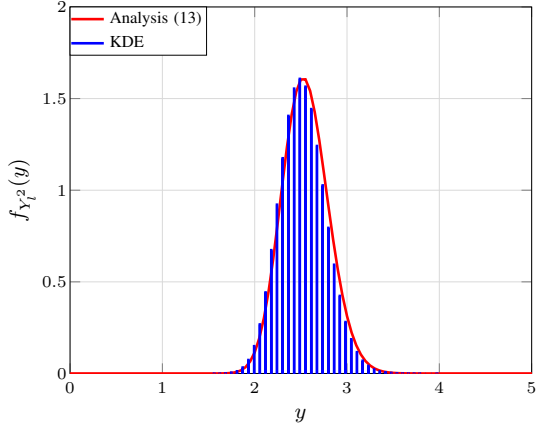
\begin{figure}[t]
\centering
\scalebox{\sca}{\begin{tikzpicture}
\begin{axis}[
        xmin=0, xmax=5,        ymin=0, ymax=2,
    xlabel={\xyfont $y$},       ylabel={\xyfont $f_{Y_{l}^2}(y)$},
    xlabel style={black, font=\fontfamily{cmr}\footnotesize , yshift=0mm},
    ylabel style={black, font=\fontfamily{cmr}\footnotesize, yshift=-1.5mm},
    y tick label style={black, font=\fontfamily{ppl}\tickfont , xshift=0.75mm},
    x tick label style={black, font=\fontfamily{ppl}\tickfont , yshift=0mm},
 grid=both, \majorgrid,  \minorgrid,
legend style={at={(axis cs:0,2)},anchor=north west},
 legend style={font=
\fontsize{7pt}{2pt}\selectfont,inner sep=0pt,outer sep=0pt},legend image post style={scale=0.8},  
               width=8.4cm,             height=7cm,  
legend cell align={left}, legend entries={Analysis (\ref{finalpdfnew}),KDE,},
 ]

\color{black}
\usetikzlibrary{shapes}

\addplot[line width=1.1pt,color=red,solid] 
coordinates{ (-50, 1)};

\addplot[line width=1.1pt, color=blue,fill=blue] 
coordinates{ (-50, 1)};

\addplot[line width=1.1pt,color=red, mark=None, mark size=2.5, mark options=solid,solid] 
   coordinates{(0.0, 0.0)(0.05, 6.067811156570168e-134)(0.1, 2.0905744179408494e-103)(0.15000000000000002, 6.468112983340815e-86)(0.2, 9.113597559600905e-74)(0.25, 1.5195114317664612e-64)(0.30000000000000004, 3.567733879457494e-57)(0.35000000000000003, 4.420774938412589e-51)(0.4, 6.36055982496139e-46)(0.45, 1.7671999791134996e-41)(0.5, 1.3418380606762483e-37)(0.55, 3.567138065369387e-34)(0.6000000000000001, 3.986384878530962e-31)(0.65, 2.1517962891110733e-28)(0.7000000000000001, 6.2499395894018e-26)(0.75, 1.0640808589587852e-23)(0.8, 1.137793361540973e-21)(0.8500000000000001, 8.084482127709766e-20)(0.9, 3.99985632048574e-18)(0.9500000000000001, 1.4330038092373483e-16)(1.0, 3.842814666417584e-15)(1.05, 7.934720324984983e-14)(1.1, 1.2925873220057961e-12)(1.1500000000000001, 1.6966428919129005e-11)(1.2000000000000002, 1.8277207379197555e-10)(1.25, 1.64220028553076e-09)(1.3, 1.2483098767483184e-08)(1.35, 8.12983019923848e-08)(1.4000000000000001, 4.587622898582354e-07)(1.4500000000000002, 2.2657455316142783e-06)(1.5, 9.88273529244732e-06)(1.55, 3.838176816975545e-05)(1.6, 0.0001337077458753212)(1.6500000000000001, 0.000420608703509878)(1.7000000000000002, 0.0012020873218128235)(1.75, 0.0031386701351288487)(1.8, 0.007525213786324124)(1.85, 0.016645057800348036)(1.9000000000000001, 0.034112412587708356)(1.9500000000000002, 0.06503124213941625)(2.0, 0.11574578115011332)(2.0500000000000003, 0.192990060350967)(2.1, 0.3023973592235245)(2.15, 0.44658534833848)(2.2, 0.6232996774397631)(2.25, 0.8242483476369453)(2.3000000000000003, 1.0351852896959925)(2.35, 1.237486664786033)(2.4000000000000004, 1.4110036240235995)(2.45, 1.5375448245484027)(2.5, 1.6041133483284111)(2.5500000000000003, 1.6050856736564243)(2.6, 1.5428438577915986)(2.6500000000000004, 1.426825114244789)(2.7, 1.2713691935748186)(2.75, 1.092993110269035)(2.8000000000000003, 0.9077548134313216)(2.85, 0.7292155430630708)(2.9000000000000004, 0.5672606007759141)(2.95, 0.42778543519306766)(3.0, 0.3130684414211904)(3.0500000000000003, 0.2225631647273411)(3.1, 0.1538430019807096)(3.1500000000000004, 0.10349104086173612)(3.2, 0.06781097283290974)(3.25, 0.043313549741929414)(3.3000000000000003, 0.026990585303453125)(3.35, 0.016420577761097236)(3.4000000000000004, 0.009760244651648507)(3.45, 0.005671842101972072)(3.5, 0.00322448975876863)(3.5500000000000003, 0.0017944922385110934)(3.6, 0.0009781935820586188)(3.6500000000000004, 0.0005225873482038021)(3.7, 0.00027376751505533135)(3.75, 0.00014070916063238577)(3.8000000000000003, 7.099040959218463e-05)(3.85, 3.517423437207276e-05)(3.9000000000000004, 1.712379165962801e-05)(3.95, 8.194461763747332e-06)(4.0, 3.856325603781494e-06)(4.05, 1.7854195380289232e-06)(4.1000000000000005, 8.135684666331995e-07)(4.15, 3.650083142877254e-07)(4.2, 1.6129756059579354e-07)(4.25, 7.023038748586257e-08)(4.3, 3.014011571820157e-08)(4.3500000000000005, 1.2753591407593316e-08)(4.4, 5.322650948530732e-09)(4.45, 2.191627781340104e-09)(4.5, 8.905940914345344e-10)(4.55, 3.5726754463541107e-10)(4.6000000000000005, 1.4152411767621626e-10)(4.65, 5.5374383473805244e-11)(4.7, 2.1406394828356585e-11)(4.75, 8.1779664042092e-12)(4.800000000000001, 3.088316293821758e-12)(4.8500000000000005, 1.1531267273318848e-12)(4.9, 4.2580628659928085e-13)(4.95, 1.5553378471495418e-13)(5.0, 5.620953282970745e-14)(5.050000000000001, 2.0102953431910612e-14)(5.1000000000000005, 7.116456208124947e-15)(5.15, 2.4940690464358263e-15)(5.2, 8.655221337118776e-16)(5.25, 2.974779082263496e-16)(5.300000000000001, 1.012785761895837e-16)(5.3500000000000005, 3.416205888570352e-17)(5.4, 1.1418502883606336e-17)(5.45, 3.782561478610351e-18)(5.5, 1.2420684952107837e-18)(5.550000000000001, 4.0434950010319026e-19)(5.6000000000000005, 1.3052308256916e-19)(5.65, 4.178325034386754e-20)(5.7, 1.326677361072691e-20)(5.75, 4.178675580252865e-21)(5.800000000000001, 1.3058180323751016e-21)(5.8500000000000005, 4.0490786374278055e-22)(5.9, 1.2459960766184384e-22)(5.95, 3.805572237905847e-23)};

\addplot[ybar, bar width=0.1pt,line width=1pt,color=blue] 
   coordinates{(1.5606390048960244, 6.463839040369705e-05)(1.6225217339525204, 0.0002747131592157135)(1.6844044630090163, 0.0016321193576933509)(1.7462871920655123, 0.005090273244291162)(1.8081699211220081, 0.01253984773831723)(1.8700526501785042, 0.033062536691491166)(1.931935379235, 0.07407559540263682)(1.993818108291496, 0.1494924374061504)(2.055700837347992, 0.2682170009801419)(2.117583566404488, 0.4427568146677224)(2.1794662954609842, 0.6734189108233194)(2.24134902451748, 0.9216464895711179)(2.303231753573976, 1.1749643415632076)(2.3651144826304717, 1.405901150878007)(2.426997211686968, 1.5547148851849293)(2.488879940743464, 1.6079930784751766)(2.5507626697999597, 1.565380219601528)(2.612645398856456, 1.4431005445553449)(2.674528127912952, 1.241881235228635)(2.7364108569694476, 1.0267646719651307)(2.7982935860259435, 0.7949714039774721)(2.8601763150824393, 0.5937197754555563)(2.9220590441389356, 0.4236723299010309)(2.983941773195432, 0.2810477214752758)(3.0458245022519277, 0.18706350182829998)(3.1077072313084235, 0.116591496690669)(3.1695899603649194, 0.07026193036881895)(3.231472689421415, 0.041853357786393694)(3.2933554184779115, 0.02226792549407372)(3.3552381475344073, 0.011651069870266354)(3.4171208765909036, 0.007013265358801157)(3.4790036056473994, 0.003555111472203351)(3.5408863347038952, 0.0015836405648905836)(3.602769063760391, 0.0007433414896425189)(3.664651792816887, 0.0003878303424221838)(3.7265345218733827, 0.00012927678080739273)(3.7884172509298795, 9.695758560554595e-05)(3.8502999799863753, 0.0)(3.912182709042871, 0.0)(3.974065438099367, 1.6159597600924325e-05)};

\end{axis}
\end{tikzpicture}}
\vspace{-2.0mm}
\renewcommand{\figurename}{\footnotesize Fig.}
\captionwidth
\caption{\footnotesize PDF validation of $\snr_{\U}$ in (\ref{finalpdfnew}) using kernel density estimation (KDE) method for $LM=120$, $L=3$, $d_{\ris}=100$ m 
$d_{\U}=30$ m, $\kappa_h=\kappa_g=10$, the rest of the parameters are provided in Table\ref{table1} in the numerical section.} 
\label{PDF_verif}
\vspace{-3mm}
\end{figure}

{\bf Lemma 1.} \emph{For a multi-sector BD-RIS aided multi-user network operating over Rician fading channels, the CDF, PDF and MGF of $\snr_{\U}$ can be, respectively, given as follows}:\vspace{0mm}
\begin{align}\label{finalcdfnew}
F_{\snr_{\U}}(y) =\frac{1}{\Gamma(k_{\ky} )}\gamma\qty(k_{\ky},\frac{y}{\alpha_{\U}\rho_{\U}\theta_{\ky}}), 
\end{align}
\vspace{0mm}
\begin{align}\label{finalpdfnew}
f_{\snr_{\U}}(y) = \frac{ y^{k_{\ky}-1} e^{-\frac{y}{\alpha_{\U}\rho_{\U}\theta}}}{\Gamma(k_{\ky}) \qty(\alpha_{\U}\rho_{\U}\theta_{\ky})^{k_{\ky}}}, 
\end{align}
\emph{and}
\vspace{2mm}
\begin{equation}\label{gammaMGF}
M_{\gamma_{\U}}(s) = \qty(1-s\mkern1mu\rho_{\U}\alpha_{\U}\theta_{\ky})^{-k_{\ky}}
\end{equation}
\emph{where $\Gamma(.)$ is the Gamma function, 
$\gamma(\cdot, \cdot)$ is the lower incomplete gamma function \cite{gradshteyn1988tables},}
\begin{align}\label{kys}
 k_{\ky} = \frac{\mkern1.5mu\mathbb E [Y^2_{k_l}]^2}{\mathbb{V}\mathrm{ar}[Y^2_{k_l}]} 
 \end{align}
\begin{align}
 \theta_{\ky} = \frac{\mathbb{V}\mathrm{ar}[Y^2_{k_l}]}{\mathbb E [Y^2_{k_l}]}, 
\end{align}
\begin{align} 
\e[Y^2_{k_l}]=&\frac{\Gamma(k_{Y_{k_l}}+2)}{\Gamma(k_{Y_{k_l}})}\theta_{Y_{k_l}}^2,
\end{align}
\begin{align} 
\mathbb{V}\mathrm{ar}[Y^2_{k_l}]=\qty(\frac{\Gamma(k_{Y_{k_l}}+4)}{\Gamma(k_{Y_{k_l}})}-\frac{\Gamma(k_{Y_{k_l}}+2)^2}{\Gamma(k_{Y_{k_l}})^2})\theta_
{Y_{k_l}}^4
\end{align}
\begin{align}
k_{Y_{k_l}} = \frac{\e[Y_{k_l}]^2}{\mathbb{V}\mathrm{ar}[Y_{k_l}]}
\end{align}
\begin{align}
\theta_{Y_{k_l}}= \frac{\mathbb{V}\mathrm{ar}[Y_{k_l}]}{\e[Y_{k_l}]}, 
\end{align}
\begin{align}\label{meannn}
   \e\qty[Y_{k_l}]= \frac{\ml\pi L_{\frac{1}{2}}(-\kappa_h)L_{\frac{1}{2}}(-\kappa_g)}{4\sqrt{(\kappa_h+1)(\kappa_g+1)}},
\end{align}
\begin{align}\label{varrr}
    \mathbb V \mathrm{ar}\qty[Y_{k_l}] =  \ml -
   \frac{\ml\pi^2 L_{\frac{1}{2}}(-\kappa_h)^2L_{\frac{1}{2}}(-\kappa_g)^2}{16(\kappa_h + 1)(\kappa_g + 1)}.
\end{align}
\textbf{\textit{Proof}}: Please refer to Appendix \customref{AA}{A}. \hfill $\blacksquare$\\
A numerical validation through kernel density estimation (KDE) method for the obtained PDF of $\snr_{\U}$ is shown in Fig. \ref{PDF_verif}, where the simulation and
the analytical results match perfectly. Moreover, the accuracy of the CDF in (\ref{finalcdfnew}), and the MGF in (\ref{gammaMGF}) will be also validated by the
performance curves in the numerical results presented in Section \ref{sec:numerical}.

\subsection{Spectral Efficiency}
The achievable spectral efficiency can be obtained as
\begin{align}\label{ach1}
\mathcal{R}= \mathbb E\qty[\frac{\Lo}{\LL}\sum_{l=1}^{L} \sum_{k=1}^{K_{l}}\log_2 \qty(1 +\rho_{\U}\alpha_{\U} Y_{k_l}^2)].
\end{align}

By using Jensen’s inequality, a lower bound of the spectral efficiency  in \eqref{ach1} is given by
\begin{align}
\mathcal{R} \geq \mathcal{\Tilde{R}} = \frac{\Lo}{\LL}\sum_{l=1}^{L} \sum_{k=1}^{K_{l}}\log_2 \qty(1 + \e\qty[\rho_{\U}\alpha_{\U} Y_{k_l}^2]).
\end{align}
which can be obtained in closed-form as
\begin{align}\label{ach}
\mathcal{R}= \frac{\Lo}{\LL}\sum_{l=1}^{L} \sum_{k=1}^{K_{l}}\log_2 \qty(1 + \frac{\rho_{\U}\alpha_{\U}\Gamma(k_{Y_{k_l}}+2)}{\Gamma(k_{Y_{k_l}})}\theta_{Y_{k_l}}^2).
\end{align}
\textbf{Proposition 1.} \emph{For a fixed number of configurable elements, the impact of beam directivity outweighs that of BD-RIS sector dimensions. For instance, a 6-sectors BD-RIS provides higher spectral efficiency than that of a 2-sectors setup, where the expressions of spectral efficiency for 2- and 6-sectors BD-RIS are respectively given as:}
\begin{align}\label{se2}
\mathcal{R}=\frac{\Lo}{\LL}\sum_{l=1}^{L} \sum_{k=1}^{K_{l}}\log_2 \qty(1 + A\qty[ 0.25Z^2\Omega+0.5Z\qty(1-\Omega) ]),
\end{align}
\begin{align}\label{se6}
\mathcal{R}=\frac{\Lo}{\LL}\sum_{l=1}^{L} \sum_{k=1}^{K_{l}}\log_2 \qty(1 + A\qty[ 1.54Z^2\Omega+9.28Z\qty(1-\Omega)]),
\end{align}
\emph{where $Z$ is the total number of configurable elements for the whole BD-RIS, i.e., $Z=LM$, and,}
\begin{align}
\Omega=\common.
\end{align}

\textbf{\textit{Proof}}: Please refer to Appendix \customref{BBB}{B}. \hfill $\blacksquare$\\
\subsection{Outage Probability}\label{sec:outage_and_problem}
Outage probability represents the likelihood that a user link fails to meet a certain predefined quality-of-service (QoS) requirement. In this system model, the QoS requirement is specified in terms of a minimum SNR that is given in \eqref{eq:user_SNR}.
Thus, a user's outage probability can be expressed as
\begin{align}\label{outdef2}
\p_{\U}(\mathcal{O})& = \Pr\qty[\snr_{\U}\leq \psi_{\U}]\nonumber\\
&=F_{\snr_{\U}}( \psi_{\U}),
\end{align}
where $\psi_{\U}=2^{\LFR_{\U}}-1$.
Substituting the CDF in (\ref{finalcdfnew}) into (\ref{outdef2}), the outage probability closed-form expression can be given as 
\begin{equation}\label{outcdf}
\p_{\U}(\mathcal{O}) = \frac{\gamma\qty(Mk,\frac{\sqrt{\psi_{\U}}}{\theta})}{\Gamma(Mk)}.
\end{equation}
\textbf{Corollary 1.} \emph{The outage performance of a 6-sector BD-RIS is expected to surpass that of a 2-sector BD-RIS. This inference stems from the observation that increasing the number of sectors results in improved achievable spectral efficiency.}
\subsection{Diversity Order}
The diversity order represents the rate at which the outage probability  decreases as the SNR increases. It quantifies the system resilience to fading, indicating how effectively the system can exploit diversity techniques to combat channel impairments. A higher diversity order implies faster outage probability decay, indicating better performance in challenging communication environments. Specifically, the diversity order represents the negative slope of the outage probability as the SNR approaches infinity \cite{diversitydef}.
Moreover, the outage probability can be asymptotically represented as $ \p_{\U} \approx (G_c \rho_{\U})^{-G_d}$,
where $ G_d $ is the diversity order and $ G_c$ is a measure of the coding gain \cite{diversitydef}.
The analytical expression for $\p_{\U}(\mathcal{O})$ can be given by 
\begin{align}\label{asymtotic_closed}
\p^{\infty}_{\U}(\mathcal{O}) =\qty[  \frac{\alpha_{\U}\theta_{\ky}\rho_{\U}}{ \psi_{\U}\Gamma\qty(k_{\ky}+1)^{\frac{-1}{k_{\ky}}} } ] ^{-k_{\ky}},
\end{align}
where the coding gain is $\scalemath{0.9}{G_c = \qty[  \frac{\alpha_{\U}\theta_{\ky}}{\psi_{\U}\Gamma\qty(k_{\ky}+1)^{\frac{-1}{k_{\ky}}} } ]}$, and the diversity order is $G_d=k_{\ky}$. 

\textbf{\textit{Proof}}: The lower incomplete gamma function can be expanded as follows \cite[Eq. (8.354.1)]{gradshteyn1988tables}
\begin{align}\label{sum}
\gamma(k_{\ky},\rho_{\U}\alpha_{\U}\theta_{\ky}) = \sum_{n=0}^{\infty} \frac{(-1)^n \left(\frac{y}{\rho_{\U}\alpha_{\U}\theta_{\ky}}\right)^{k_{\ky} + n}}{n! (k_{\ky} + n)}    
\end{align}
At high SNR, $\rho_{\U}$ approaches $\infty$.
Therefore, the first term in the summation in (\ref{sum}) is the dominate one,
and by considering this, we have:
\begin{align}
\gamma(k_{\ky},\rho_{\U}\alpha_{\U}\theta_{\ky}) =  \frac{\left(\frac{y}{\rho_{\U}\alpha_{\U}\theta_{\ky}}\right)^{k_{\ky}}}{ k_{\ky} \Gamma\qty(k_{\ky})},    
\end{align}
substituting in (\ref{outcdf}) and applying some algebraic manipulation we obtain (\ref{asymtotic_closed}).\hfill $\blacksquare$\\
\textbf{Remark 1.} \emph{The diversity order $k_{\ky}$ is dependent on various factors including the number of configurable elements per sector $\ml$, and the Rician channel factors $\kappa_h$ and $\kappa_g$, i.e. the LoS connection establishment through the deployment of the BD-RIS. These parameters collectively influence the system ability to achieve diversity gain, highlighting the significance of their careful consideration in the design and deployment of the BD-RIS-aided networks.}
%
%


\noindent
\textbf{Remark 2.} \emph{ Although designing a multi-sector BD-RIS with a high number of sectors achieves better outage performance than with a low number of sectors for a given number of configurable elements, it becomes evident that a higher diversity order is obtained with a low number of sectors compared to a high number. This is because the diversity order is dependent on the number of configurable elements per sector.}

\subsection{Symbol Error Probability}
In this subsection, we analyze  the SEP, a critical metric for evaluating the system performance. SEP quantifies the likelihood of erroneous detection or reception of symbols in digital communication systems. It represents the probability that a transmitted symbol is incorrectly decoded or received as a different symbol due to noise or other impairments in the channel. In our system model, we leverage the MGF-based approach to compute the average SEP specifically tailored for $M$-phase shift keying (PSK) signaling, which is obtained as \cite[Eq. (5.67)]{digitalcommun2}:
%
\begin{equation}\label{M-ary-pe}
P_s = \frac{1}{\pi} \int_0^{(M-1)\pi/M} 
M_{\gamma_{\U}} \qty(-\frac{2\sin^2\qty(\pi/M)}{2\sin^2\delta}) d\delta,
\end{equation}
\textbf{Proposition 2.} \emph{By evaluating for BPSK $(M=2)$ as a representative example of $M-$ary modulation scheme, and substituting the MGF of (\ref{gammaMGF}) into  (\ref{M-ary-pe}) we have
\begin{equation}
P_s = \frac{1}{\pi} \int_0^{\pi/2} 
\qty(1+\frac{\rho_{\U}\alpha_{\U}\theta_{\ky}}{\sin^2\delta})^{-k_{\ky}} d\delta,
\end{equation}
with the help of \cite[Eq. (5.17b)]{digitalcommun2}, we can solve the above integral, and thus, the closed-form expression for the average SEP of BPSK can be given as
 \begin{align}\label{cfsep}
P_s=&
\frac{1}{2\sqrt{\pi}}\frac{\sqrt{\rho_{\U}\alpha_{\U}\theta_{\ky}}}{\qty(1+\rho_{\U}\alpha_{\U}\theta_{\ky})^{k_{\ky}+\frac{1}{2}}}
\frac{\Gamma\qty(k_{\ky}+\frac{1}{2})}{\Gamma\qty(k_{\ky}+1)}\nonumber\\
&\times{}_2F_1\qty(1,k_{\ky}+\frac{1}{2}; k_{\ky}+1, \frac{1}{1+\rho_{\U}\alpha_{\U}\theta_{\ky}}),
\end{align}
where $_2F_1\qty(.,. ; ., .)$ is the Gauss hypergeometric function.}

Next, we analyze the asymptotic SEP in high-SNR regimes, facilitating an insightful investigation on the system performance under favorable conditions. Leveraging the property $_2F_1\qty(.,. ; ., 0) =1$ \cite{2f1} at high SNR, along with some algebraic manipulations, we can obtain the asymptotic SEP closed-form expression as follows:
%
\begin{align}\label{cfsep}
P_s=
\qty(\alpha_{\U}\theta_{\ky}
\qty(\frac{2\sqrt{\pi}\Gamma\qty(k_{\ky}+1)}{\Gamma\qty(k_{\ky}+\frac{1}{2})})^{\frac{1}{k_{\ky}}}\rho_{\U})^{-k_{\ky}}
\end{align}
where the coding gain is $\scalemath{0.9}{G_c = \qty[\alpha_{\U}\theta_{\ky}
\qty(\frac{2\sqrt{\pi}\Gamma\qty(k_{\ky}+1)}{\Gamma\qty(k_{\ky}+\frac{1}{2})})^{\frac{1}{k_{\ky}}}]}$,
and the diversity order is $G_d=k_{\ky}$.

\noindent
\textbf{Remark 3.} \emph{The asymptotic SEP exhibits similar behavior to the asymptotic outage probability, given their shared diversity order of $k_{\ky}$, which emphasizes the interrelation between these key performance metrics in evaluating system reliability and robustness.}

\subsection{Energy Efficiency}
Energy efficiency is an important metric for evaluating system performance. In this subsection, we introduce the definition and derivation of the energy efficiency for the studied system model, which is measured in bits-per-joule (b/J) and defined as 
\begin{equation}
EE = W \frac{\mathcal{R}}{\mathcal{P}_{\mathrm{total}}},
\end{equation}
where $W$ denotes the transmission bandwidth in Hz, $\mathcal{R}$ is the spectral efficiency as defined in \eqref{ach}, and ${\mathcal{P}_{\mathrm{total}}}$ is the total power consumed by the BD-RIS-aided multi-user system. 

To compute the total power consumption ${\mathcal{P}_{\mathrm{total}}}$ in this system, we provide a description of the energy consumption model. The network energy consumption includes the BS transmit power, alongside the hardware static power consumed by the BS, user terminals, and the BD-RIS. Moreover, due to the passive nature of the meta-surfaces in a BD-RIS, the absence of amplifiers results in minimal energy consumption. It is important to highlight that no power is directly utilized by the BD-RIS as its elements are passive and do not actively alter the incoming signal strength. In fact, any enhancement by the BD-RIS to the signal occurs through adjusting the phase shifts of its elements to maximize the cascaded channels. Hence, the total power consumption between the BS and $U_{\U}$ can be expressed as \cite{p_cons_model}:
\begin{equation}\label{pmodel}
\mathcal{P}_{\U} =  p_{\U}/\nu + P_{\mathrm{UE},\U} + P_{\mathrm{BS}} + P_{\mathrm{RIS}}, 
\end{equation}
where $P_{\mathrm{UE},\U}$, and $P_{\mathrm{BS}}$ denote the hardware static power dissipated by $U_{\U}$ user equipment, and total hardware static power consumption at the BS, respectively. In \eqref{pmodel}, $\nu \in (0, 1]$ accounts for the power amplifier efficiency \cite{Bjornson2020}.
Additionally, we make the assumption that the transmit amplifier operates within its linear range, and its circuit power remains independent of the communication rate \cite{p_cons_model}. These assumptions align with standard wireless communication systems, where amplifiers are engineered to operate within their linear transfer function region, and where the hardware-dissipated power can be represented by a constant offset \cite{p_cons_model}.
%

The power consumption of the RIS arises from two primary sources. First, its individual configurable elements, which perform phase shifting in the incoming signal. Additionally, with the adoption of the TS protocol, wherein only the sector corresponding to each $U_{\U}$ is activated, the power dissipation at the BD-RIS during each user's orthogonal time slot is given by $MP_e$, where $P_e$ represents the power consumption of the RIS phase shifter. Consequently, the total power dissipation at the BD-RIS for all $K$ user transmissions amounts to $P_{\mathrm{RIS}}=KMP_e$. Secondly, there is $P_{sw}$, which accounts for the power dissipation caused by the circuitry involved in switching between the different sectors required for TS employment.
Hence, the total power consumption for operating the BD-RIS-assisted multi-user system can be expressed as:
\begin{equation}\label{pmodelequ}
\mathcal{P}_{\mathrm{total}} = \sum_{l=1}^{L}\sum_{k=1}^{K_l} \qty( p_{\U}/\nu + P_{\mathrm{UE},\U}) + P_{\mathrm{BS}} + KMP_e+P_{sw}.
\end{equation}
Therefore, based on the derived spectral efficiency in \eqref{ach}, the overall energy efficiency can be expressed as follows:
\begin{equation}
EE = W \frac{\frac{1}{K}\sum\limits_{l=1}^{L}\sum\limits_{k=1}^{K_l}\log_2 \qty(1 + \frac{\rho_{\U}\alpha_{\U}\Gamma(k_{Y_{k_l}}+2)}{\Gamma(k_{Y_{k_l}})}\theta_{Y_{k_l}}^2)}
{\sum\limits_{l=1}^{L}\sum\limits_{k=1}^{K_l} ( p_{\U}/\nu +K P_{\mathrm{UE},k} ) + P_{\mathrm{BS}} + KMP_e+P_{sw}}.
\end{equation}

\section{Numerical Results}\label{sec:numerical}
In this section, we conduct simulations to validate the performance of the proposed system. Specifically, we numerically evaluate the derived outage probability, average achievable rate, and diversity order analytical expressions.  To ensure precision, Monte Carlo simulations are performed with a total of $10^7$ random Rician fading channel realizations. The targeted data rates per user $R_{\U}$ are measured in bits per channel use (BPCU). Unless explicitly specified, the simulation parameters are detailed in Table \ref{table1}.

\begin{small}
\begin{table}[ht]
\caption{\vspace{0.5mm}Simulation Parameters}\vspace{-2mm}
\label{table1}
\centering
\renewcommand{\arraystretch}{1.3} 
\begin{tabular}{|c|c|}
\hline
\bf Parameter & \bf Value  \\
\hline
Transmit signal frequency $f$ & $2.4$ GHz \\ 
$W$ & $10$ MHz \\ 
$\nu$ & $0.5$\\ 
$P_{\mathrm{UE},k}= P_{\mathrm{BS}} =  P_{sw}$  & $10$ dBm  \\ 
Bandwidth BW: & $10$ MHz \\ 
$P_e$ & $0.5$ mW \\ 
$K$ & $6$   \\
$d_{\ris}$ &  $100$ m \\
$d_{\U}$ & $30$ m \\
$\eta$ for Rician &  $2$   \\
$\eta$ for Rayleigh  &  $3$   \\
$G_t=G_r$  &  $1$   \\
Noise power $\sigma_{n_{\UU}}^2$ &$-80$ dBm  \\
\hline
\end{tabular}
\end{table}
\end{small}


\subsection{Spectral Efficiency}
\begin{figure}[t]
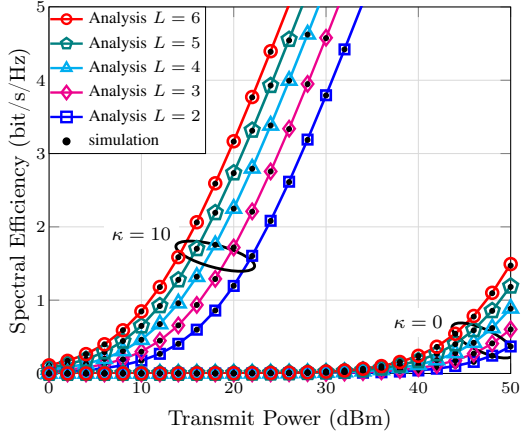

\centering
\scalebox{\sca}{
}
\vspace{-1.0mm}
\renewcommand{\figurename}{\footnotesize Fig.}
\captionwidth
\caption{\footnotesize  Spectral efficiency versus transmit power $p_{\U}$ in dBm for different number of sectors with  $L\ml=360$ and $\kappa_h=\kappa_g=\kappa$}
\label{se}
\vspace{-1mm}
\end{figure}

In Fig. \ref{se}, we investigate the spectral efficiency of the multi-sector BD-RIS assisted communication for finite number of configurable elements, set at $L\ml=360$, distributed across various sector counts. As observed, a BD-RIS with $L=6$ sectors shows a significant gain compared to a BD-RIS with $L=2$ sectors. For example, in the case of Rician fading, achieving a spectral efficiency of 2 bit/sec/Hz demands a transmit power of $p=16$ dBm for the 6-sector BD-RIS, whereas the same spectral efficiency requires $p_{\U}=24$ dBm for the 2-sector configuration, serving the same total number of users. This disparity can be attributed to the narrowing beamwidth of each BD-RIS element with increasing sector count, leading to higher directionality. Specifically, transitioning from 2 sectors to 6 sectors yields an average spectral efficiency gain of $182\%$ for $\kappa=10$ and $14\%$ for $\kappa=0$.
These findings highlight the substantial spectral efficiency gains achieved through sectorization in multi-sector BD-RIS systems, thereby maximizing channel capacity utilization, especially in Rician fading scenarios.


In Fig. \ref{ele-rate}, the spectral efficiency versus the total number of the whole BD-RIS elements is compared under different number of sectors from $L=2$ to $L=6$. A similar trend is seen as in Fig. \ref{out_ele}, in which one can confirm that increasing the number of sectors has a significant effect on reducing the number of elements required to achieve a certain spectral efficiency. 
For instance, to realize a targeted spectral efficiency $\bar{\mathcal{R}}=6$ bit/sec/Hz, the number of elements required for a 6-sectors BD-RIS is $L\ml=330$, while the number of elements needed to achieve the same spectral efficiency increases to $L\ml=800$ for a 2-sector BD-RIS.

\subsection{Outage probability}
In Fig. \ref{out_ri}, the effect of varying the number of sectors and the Rician factors $\kappa_h$ and $\kappa_g$ on the outage probability for $U_{\U}$ is investigated. Evidently, the outage performance is significantly improved as the number of sectors or the Rician factor increases, indicating enhanced system performance attributed to the multi-sector BD-RIS implementation. This also verifies Corollary 1. Moreover, the analytical expression match perfectly with simulation curves. 
One can also deduce that the outage performance depends on the precise positioning of the multi-sector BD-RIS to establish virtual LoS connections between the BS and  users, where this strategic placement ensures minimizing the probability of outage events. 

In Fig. \ref{out_ele}, we can observe that increasing the number of sectors notably reduces the required number of elements to attain a certain outage performance level. In particular, the multi-sector BD-RIS outperforms the STAR-RIS configuration by minimizing the number of elements needed to realize a given outage value. 
For instance, achieving a $10^{-2}$ outage probability with $L = 6$, the multi-sector setup requires approximately $M \approx 350$ elements compared to $M \approx 825$ in the STAR-RIS setup, marking a $135\%$ reduction in required elements through sectorization of the BD-RIS. The findings emphasize the efficiency of multi-sector BD-RIS deployments in achieving outage performance targets with fewer elements compared to traditional STAR-RIS setups. This highlights the potential for significant resource savings and improved system scalability, offering a compelling advantage for practical implementation in communication networks.


\subsection{Diversity order}
In Fig. \ref{asym_LM}, we investigate the system outage performance across varying numbers of $L\ml$ configurable elements and different sector counts. Clearly, the derived analytical and asymptotic expressions align closely with simulation curves, validating their accuracy in characterizing the system outage behavior. Moreover, the observed increase in diversity gain with higher numbers of elements per sector corroborates the analysis of diversity order and aligns with the insights from Remark 1 and Remark 2. 
Specifically, the outage probability curves exhibit a steeper slope for lower numbers of sectors, attributed to the relationship between diversity orders and the configurable elements per sector $\ml$. However, this superiority of fewer sectors is observed primarily below an outage probability of $10^{-5}$, which remains below practical outage probability values typically encountered in real systems ($10^{-4}$ to $10^{-5}$). Further, increasing $L\ml$ yields a notable enhancement in the outage curves, diminishing the superiority in diversity order for a low number of sectors as $L\ml$ increases.

 
In Fig. \ref{asym_RI}, the outage and asymptotic behavior are investigated versus transmit power for different Rician factors $\kappa_h$ and $\kappa_g$. Evidently, the outage performance is enhanced as the Rician factor increases, which is attributed to the presence of stronger LoS components in the communication channel, implying that the outage performance depends on the precise placement of the multi-sector BD-RIS to establish virtual LoS connections between the BS and users. This observation also validate the diversity order analysis since $G_d$, is a function of the Rician factors $\kappa_h$ and $\kappa_g$. Furthermore, the correlation between Rician factors and the diversity order highlights their influence on system reliability, affirming the importance of accounting for channel characteristics in the design and deployment of multi-sector BD-RIS setups. 


\begin{figure}[t]
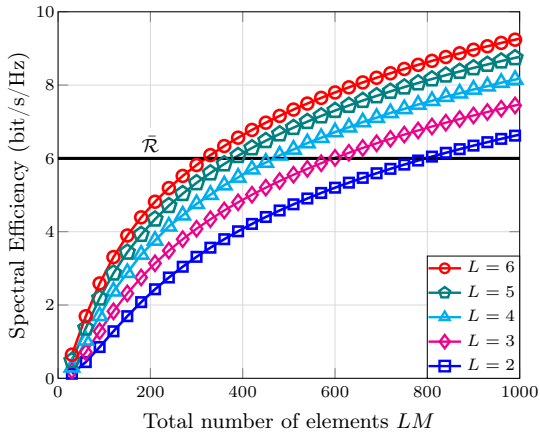

\centering
\scalebox{\sca}{%
}
\vspace{-1.0mm}
\renewcommand{\figurename}{\footnotesize Fig.}
\captionwidth
\caption{\footnotesize Spectral efficiency vs the total number of elements for different number of sectors while fixing $LM=360$.}
\label{ele-rate}
\vspace{-1mm}
\end{figure}


\begin{figure}[t]
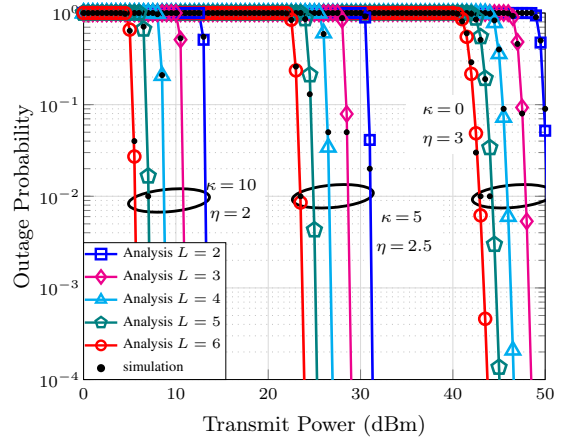

\centering
\scalebox{\sca}{%
}
\vspace{-1.0mm}
\renewcommand{\figurename}{\footnotesize Fig.}
\captionwidth
\caption{\footnotesize Outage probability versus $p_k$ in dBm for different number of sectors with $LM=960$ and $\kappa_h=\kappa_g=\kappa$ for $R_{\U} = 0.25$.}
\label{out_ri}
\vspace{-1mm}
\end{figure}
 \vspace{0mm}

\begin{figure}[t]
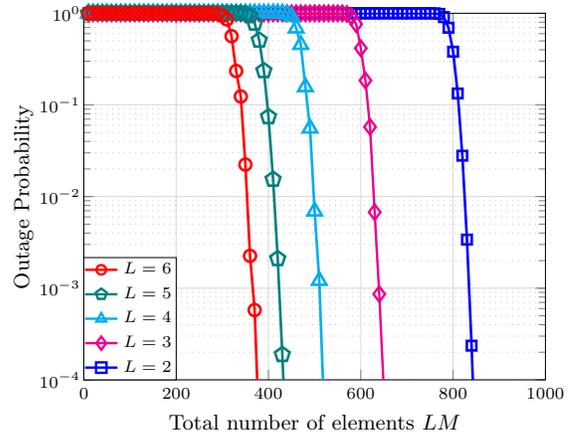

\centering
\scalebox{\sca}{%
}
\vspace{-1.0mm}
\renewcommand{\figurename}{\footnotesize Fig.}
\captionwidth
\caption{\footnotesize Outage probability versus total number of elements $L\ml$ for different number of sectors and $\kappa_h=\kappa_g=10$ for $R_{\U} = 0.25$.}
\label{out_ele}
\vspace{-1mm}
\end{figure}


\begin{figure}[t]
\centering
\scalebox{\sca}{%
}
\vspace{-1.0mm}
\renewcommand{\figurename}{\footnotesize Fig.}
\captionwidth
\caption{\footnotesize Asymptotic outage analysis and simulation for muli-sector BD-RIS against transmit power for different number of sectors and different $LM$ with $\kappa_h=\kappa_g=10$, $d_{\ris}=30$ m and $d_{\U}=30$ m, for $R_{\U} = 0.25$.}
\label{asym_LM}
\vspace{-1mm}
\end{figure}

\begin{figure}[t]
\centering
\scalebox{\sca}{%
}
\vspace{-1.0mm}
\renewcommand{\figurename}{\footnotesize Fig.}
\captionwidth
\caption{\footnotesize Asymptotic outage analysis and simulation for muli-sector BD-RIS against transmit power for different number of sectors with  $L\ml=240$, $\kappa_h=\kappa_g=\kappa$, $d_{\ris}=30$ m and $d_{\U}=30$ m, for $R_{\U} = 0.25$.}
\label{asym_RI}
\vspace{-1mm}
\end{figure}

\begin{figure}[t]
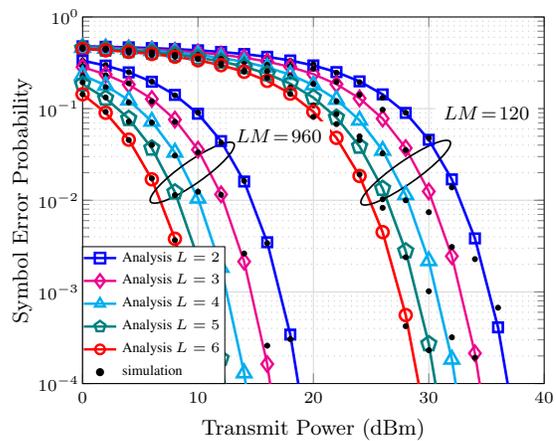

\centering
\scalebox{\sca}{%
}
\vspace{-1.0mm}
\renewcommand{\figurename}{\footnotesize Fig.}
\captionwidth
\caption{\footnotesize SEP versus $p_{\U}$ in dBm for different number of configurable elements for the whole BD-RIS and $\kappa_h=\kappa_g=10$.}
\label{sep_LM}
\vspace{-1mm}
\end{figure}

\begin{figure}[t]
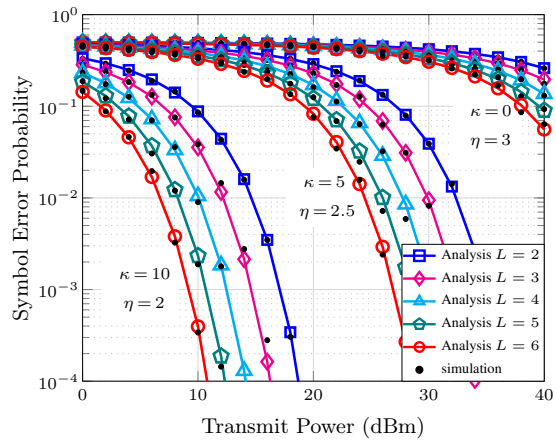

\centering
\scalebox{\sca}{%
}
\vspace{-1.0mm}
\renewcommand{\figurename}{\footnotesize Fig.}
\captionwidth
\caption{\footnotesize SEP performance for different Rician factors $\kappa_h=\kappa_g$ and different number of sectors with fixed $L\ml=960$.}
\label{sep_RICI}
\vspace{-1mm}
\end{figure}

\begin{figure}[t]
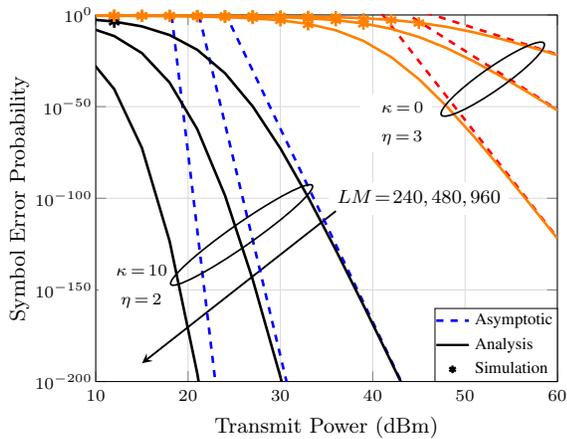

\centering
\scalebox{\sca}{%
}
\vspace{-1.0mm}
\renewcommand{\figurename}{\footnotesize Fig.}
\captionwidth
\caption{\footnotesize Asymptotic SEP analysis and simulation for different Rician factors $\kappa_g=\kappa_h=\kappa$, and different number of configurable elements for $L=6$.}
\label{sep_asym}
\vspace{-1mm}
\end{figure}

\begin{figure}[t]
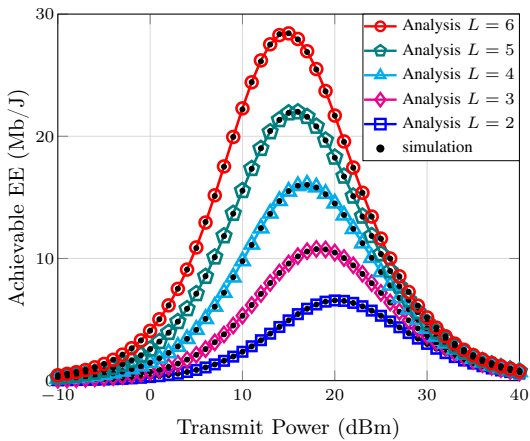

\centering
\scalebox{\sca}{%
}
\vspace{-1.0mm}
\renewcommand{\figurename}{\footnotesize Fig.}
\captionwidth
\caption{\footnotesize Energy efficiency versus transmit power $p_k$ for fixed total number of elements for the whole multi-sector $LM=360$ with different number of sectors, $P_e=0.5$ mW, and $\kappa_h=\kappa_g=10$.}
\label{eee}
\vspace{-1mm}
\end{figure}

\begin{figure}[t]
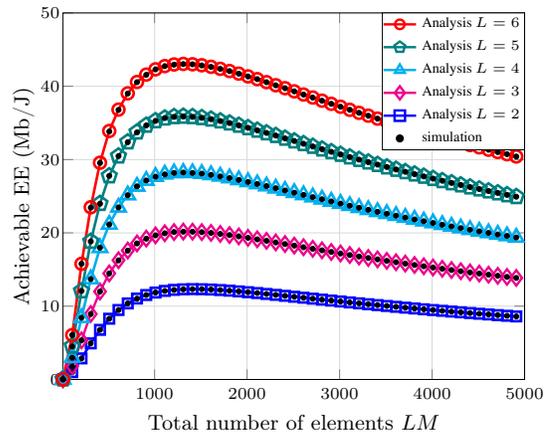

\centering
\scalebox{\sca}{%
}
\vspace{-1.0mm}
\renewcommand{\figurename}{\footnotesize Fig.}
\captionwidth
\caption{\footnotesize Energy efficiency versus the total number of elements for the whole multi-sector $LM$ with different number of sectors, $p_{\U}=17$ dBm, $P_e=0.5$ mW, and $\kappa_h=\kappa_g=10$.}
\label{elem_eff}
\vspace{-1mm}
\end{figure}

\subsection{Symbol Error Probability}

In Fig. \ref{sep_LM}, we investigate the impact of utilizing more configurable elements on the SEP performance, varying the total number of elements $L\ml$. Clearly, increasing $L\ml$ and the number of sectors enhances SEP performance. Moreover, sectoriziaton of the BD-RIS yields improved SEP performance. For example, with $LM=960$, achieving a probability value of $10^{-2}$ requires approximately $6.8$ dBm transmit power for a 6-sector BD-RIS, whereas a 2-sector BD-RIS demands roughly $14.70$ dBm, indicating a reduction in transmit power of $7.9$ dBm when transitioning from 2 to 6 sectors. This considerable reduction in the required transmit power points out the efficiency gains achieved through RIS sectorization, emphasizing its potential to significantly enhance system performance while concurrently reducing power consumption.


In Fig. \ref{sep_RICI}, we evaluate the SEP for the multi-sector BD-RIS under different fading conditions. Specifically, we vary the Rician factors and path loss exponents while keeping the total number of elements in the multi-sector BD-RIS fixed at $ML = 960$. \bla{Intuitively, it is clear from this result that an increase in the path loss exponent leads to degraded SEP performance. Additionally, SEP degrades with lower Rician factors.    However, this degradation can be mitigated by adopting solutions such as increasing transmit power and/or expanding the number of sectors in the BD-RIS. In practical systems, while increasing transmit power can provide a straightforward solution, it may lead to undesirable consequences such as increased energy consumption and interference. On the other hand, increasing the number of sectors in the BD-RIS offers a more sustainable approach, as it leverages spatial diversity to enhance communication reliability without significantly increasing energy consumption or interference levels.}



Next, we evaluate the asymptotic SEP under varying channel conditions.
Fig. \ref{sep_asym} studies the impact of the total number of configurable elements $L\ml$ and the Rician factors $\kappa_h=\kappa_g$ on the asymptotic SEP performance. As expected, a similar trend is observed as in the asymptotic outage probability analysis. Specifically, the diversity order $G_d$ increases with higher Rician factors or total number of elements $L\ml$, leading to improved diversity order, consistent with the observations for outage probability. This correlation is aligned with the insights obtained in Remark 3.


\subsection{Energy Efficiency}
In Fig. \ref{eee}, we investigate the energy efficiency performance of the multi sector BD-RIS. Clearly, when distributing the $LM_l=360$ configurable elements across $6$ sectors, significant energy efficiency enhancements are evident compared to allocating them across only $2$ sectors. Specifically, when the total number of elements $L\ml$ are divided into $6$ sectors, the maximum energy efficiency peaks at $28.87$ Mb/J with a transmit power of $15$ dBm. In contrast, when limited to $2$ sectors, the maximum efficiency diminishes to $8.52$ Mb/J at a higher transmit power of $18$ dBm. In other words, transitioning from 2 to 6 sectors yields a notable  $238\%$ increase in maximum energy efficiency, representing more than a threefold gain in system energy efficiency.

Fig. \ref{elem_eff} depicts the achievable energy efficiency plotted against the total number of configurable elements of the entire BD-RIS $LM$ for different number of sectors. It is evident that as $LM$ grows, the performance of the energy efficiency curves significantly improves owing to the enhancement in spectral efficiency. However, a further increase in $LM$ leads to a decline in the energy efficiency. This can be attributed to the fact that while the power consumption of the BD-RIS $(P_{\mathrm{RIS}})$ linearly increases with $LM$, the spectral efficiency does not increase at the same rate. Instead, it begins to saturate, as observed in Fig. \ref{ele-rate}, particularly when there is a substantial increase in the number of configurable elements. 
Furthermore, it can be clearly observed  that increasing the number of sectors can be a promising solution to mitigate the corresponding energy consumption associated with deploying a BD-RIS with a massive number of configurable elements. This is particularly relevant in scenarios where a large number of elements are required to counteract severe path loss, as the energy efficiency is degraded with increasing the number of elements. This result also implies the existence of a tradeoff between the number of configurable elements $LM$ of the BD-RIS and system energy efficiency, indicating that optimization can be achieved according to the channel conditions and usage scenarios.

In closing, the obtained results in this section highlight the importance of sectorization in practical system designs leveraging multi-sector BD-RIS technology. They demonstrate the significant impact of increasing the number of sectors on spectral and energy efficiency, emphasizing the importance of carefully considering sectorization schemes for substantial performance improvements in real-world deployments. However, it is important to mention that while increasing the number of sectors can enhance spectral and energy efficiency gains, it also introduces added hardware and control complexities. Thereby, in practical system designs, there exists a tradeoff between potential performance enhancements and the associated overhead in hardware and control intricacies.

\vspace{0mm}
\section{Conclusions}\label{sec:conc}
\vspace{0mm}
In this paper, we conducted a comprehensive performance analysis of a multi-sector BD-RIS-assisted multi-user communication system. Specifically, we derived closed-form expressions for the MGF, PDF, and CDF of the SNR per user. Furthermore, exact closed-form expressions were derived for the outage probability, achievable spectral and energy efficiency, SEP, and diversity order for the studied system model under Rician fading channels. Our high-SNR asymptotic analysis revealed that increasing the number of sectors for a fixed number of configurable elements improves outage performance at the expense of a lower diversity order. However, this convergence occurs below outage probability of $10^{-5}$. Moreover, our study provided useful insights into the nuanced performance characteristics of these systems. We found that increasing the number of sectors within the BD-RIS architecture while keeping the total number of configurable elements fixed offers a promising avenue for enhancing system performance. Numerical results demonstrated significant gains in spectral and energy efficiency when transitioning from a lower number of sectors to a higher one. These findings underscore the considerable potential of multi-sector BD-RIS configurations in optimizing spectral and energy efficiency metrics, providing valuable guidance for future system design and deployment strategies. Looking ahead, there are several promising research directions for multi-sector BD-RIS systems that can be explored for optimizing the performance and scalability of the BD-RIS architecture.


\vspace{0mm}
\section*{Appendix A}\label{AAA}
\vspace{2mm}
\section*{Derivation of $\mathbb E [Y_{\mk}]$ and  $\mathbb{V}\mathrm{ar}[Y_{\mk}]$ in (\ref{mean}) and (\ref{var}) }
The mean and variance of $\zeta_m$ and $\xi_{m,k}$ can be obtained with the help of  \cite[Eq. (59)]{MGFrician} by evaluating the $n^{th}$ moment at $n=\frac{1}{2}$ and $n=1$, respectively, as follows:
\begin{align}\label{mean1}
\mathbb E [\zeta_m]=\sqrt{\frac{\pi}{4(\kappa_h+1)}}{}_1F_1\qty(-\frac{1}{2}; 1; -\kappa_h),
\end{align}
\vspace{0mm}
\begin{align}\label{var1}
\mathbb{V}\mathrm{ar}[\zeta_m]=1-\frac{\pi}{4(\kappa_h+1)}{}_1F_1\qty(-\frac{1}{2}; 1; -\kappa_h)^{2},
\end{align}
where ${}_1F_1\qty(-\frac{1}{2}; 1; \kappa_h)$ is the confluent hypergeometric function of the first kind.
Note that, ${}_1F_1\qty(-\frac{1}{2}; 1; \kappa_h)$ can be expressed in terms of Laguerre polynomial \cite{hyperequallaguere}. 

Similarly, $\xi_{\mk}$ follows a Rician distribution with the following mean and variance:
\begin{align}\label{mean2}
\mathbb E [\xi_{\mk}]=\sqrt{\frac{\pi}{4(\kappa_g+1)}}L_{\frac{1}{2}}(-\kappa_g),
\end{align}
\vspace{0mm}
\begin{align}\label{var2}
\mathbb{V}\mathrm{ar}[\xi_{\mk}]=1-\frac{\pi}{4(\kappa_g+1)}L_{\frac{1}{2}}(-\kappa_g)^2,
\end{align}
where $L_{\frac{1}{2}}(-\kappa_g)= e^{\frac{-\kappa_g}{2}} \qty[(1 + \kappa_g) I_0(\frac{\kappa_g}{2}) + \kappa_g I_1(\frac{\kappa_g}{2})]$, $I_{\nu}(\cdot)$ is the modified $\nu$-order Bessel function of the first kind \cite[Eq. (8.431)]{gradshteyn1988tables}.
Since $\zeta_m$ and $\xi_{\mk}$ are independent random variables, the mean and variance of the their product can be determined as follows
\begin{align}\label{finalmean}
\mathbb E [Y_{\mk}]= \mathbb E [\zeta_{m}]\mathbb E [\xi_{\mk}] 
\end{align}
\vspace{-2mm}
\begin{align}\label{finalvar}
\mathbb{V}\mathrm{ar}[Y_{\mk}]=& \mathbb E [\zeta_{m}]^2\mathbb{V}\mathrm{ar}[\xi_{\mk}]+
\mathbb E [\xi_{\mk}]^2\mathbb{V}\mathrm{ar}[\zeta_{m}]\nonumber\\[3mm]
&+\mathbb{V}\mathrm{ar}[\zeta_{m}]\mathbb{V}\mathrm{ar}[\xi_{\mk}],
\end{align}
Therefore, $\e\qty[Y_{\mk}]$ and $\mathbb V \mathrm{ar}\qty[Y_{\mk}]$ can be evaluated by substituting (\ref{mean1}), (\ref{var1}), (\ref{mean2}) and (\ref{var2}), into (\ref{finalmean}) and (\ref{finalvar}).
Further, $\e\qty[Y_{k_l}]= M\e\qty[Y_{\mk}]$ and $\mathbb V \mathrm{ar}\qty[Y_{k_l}]= M\mathbb V \mathrm{ar}\qty[Y_{\mk}]$. Hence, the mean and variance of $Y_{k_l}$ can be given as follows
\begin{align}\label{mean}
   \e\qty[Y_{k_l}]= \frac{\ml\pi L_{\frac{1}{2}}(-\kappa_h)L_{\frac{1}{2}}(-\kappa_g)}{4\sqrt{(\kappa_h+1)(\kappa_g+1)}},
\end{align}
\begin{align}\label{var}
    \mathbb V \mathrm{ar}\qty[Y_{k_l}] =  \ml -
   \frac{\ml\pi^2 L_{\frac{1}{2}}(-\kappa_h)^2L_{\frac{1}{2}}(-\kappa_g)^2}{16(\kappa_h + 1)(\kappa_g + 1)},
\end{align}
and by matching the mean and variance of $Y_{k_l}$ in (\ref{mean})
and (\ref{var}), with the $k\theta$ mean and $k\theta^2$ variance of the Gamma distribution, $Y_{k_l}$ can be approximated as $Y_{k_l} \sim \Gamma(k_{Y_{k_l}}, \theta_{Y_{k_l}})$ with the shape $k_{Y_{k_l}}$ and scale $\theta_{Y_{k_l}}$ parameters:
\begin{align}\label{shapek}
 k_{Y_{k_l}} = \frac{\e[Y_{k_l}]^2}{\mathbb{V}\mathrm{ar}[Y_{k_l}]} ~~and~~  \theta_{Y_{k_l}}= \frac{\mathbb{V}\mathrm{ar}[Y_{k_l}]}{\e[Y_{k_l}]}, 
\end{align}
Accordingly, the $n^{th}$ moment of $Y_{k_l}$ can be given as 
\begin{equation} \label{momequ}
\e[Y_{l}^n]=\frac{\Gamma(k_{Y_{k_l}}+n)}{\Gamma(k_{Y_{k_l}})}\theta_{Y_{k_l}}^n,
\end{equation}
 since we are interested to approximate $Y_{k_l}^2$, we need to evaluate its mean $\e[Y^{2}_{k_l}]$ and variance $\mathbb{V}\mathrm{ar}[Y_{k_l}^2]$. Hence, the first two moments of $Y_{k_l}^2$ can be calculated from the moments of $Y_{k_l}$ by setting $n=2$ and $n=4$ in (\ref{momequ}), respectively:
\begin{align} 
\e[Y^{2}_{k_l}]=&\frac{\Gamma(k_{Y_{k_l}}+2)}{\Gamma(k_{Y_{k_l}})}\theta_{Y_{k_l}}^2,
\end{align}
\begin{align} 
\mathbb{V}\mathrm{ar}[Y_{k_l}^2]=\frac{\Gamma(k_{Y_{k_l}}+4)}{\Gamma(k_{Y_{k_l}})}\theta_
{Y_{k_l}}^4-\qty(\frac{\Gamma(k_{Y_{k_l}}+2)}{\Gamma(k_{Y_{k_l}})}\theta_{Y_{k_l}}^2)^2.
\end{align}
Now $Y^{2}_{k_l}$ can be approximated as $Y^{2}_{k_l}\sim \Gamma( k_{Y^{2}_{k_l}}, \theta_{Y^{2}_{k_l}})$ with $k_{\ky}$ and $\theta_{\ky}$ given as follows:
\begin{align}\label{shapek}
k_{\ky} = \frac{\e[Y^{2}_{k_l}]^2}{\mathbb{V}\mathrm{ar}[Y_{k_l}^2]} ~~and~~  \theta_{\ky} = \frac{\mathbb{V}\mathrm{ar}[Y_{k_l}^2]}{\e[Y^{2}_{k_l}]}, 
\end{align}
and by scaling with $\alpha_{\U}\rho_{\U}$, $\snr_{k_l}$ can be approximated as $\snr_{k_l}\sim \Gamma(k, \alpha_{\U}\rho_{\U}\theta)$, and hence, we obtain the CDF in (\ref{finalcdfnew}). Further, the PDF of (\ref{finalpdfnew}) can be obtained from \cite[Eq. (3.3.6)]{casella}, while the MGF can be given by  \cite{digitalcommun2}, and the proof is completed.
\section*{Appendix B}\label{BBB} 
\vspace{2mm}
\section*{Impact of Sectorization on Spectral Efficiency}

In order to glean more insights and to demonstrate the effect of increasing the number of sectors on the received power, let us express the spectral efficiency in terms of the mean and variance of $Y_{\lk}$, by utilizing the following relation 
\begin{align}\label{ident}
\e[Y^2_{\lk}]=\e[Y_{\lk}]^2+\mathbb{V}\mathrm{ar}[Y_{\lk}],
\end{align}
hence, by substituting (\ref{meannn}) and (\ref{varrr}) into (\ref{ident}), the spectrum efficiency in (\ref{ach}) can be rewritten as follows 
\begin{align}\label{tobeexpressed}
\mathcal{R}=\frac{1}{K}\sum_{l=1}^{L} \sum_{k=1}^{K_{l}}\log_2 \big(1 + \rho_{\U}\alpha_{\U}\qty[ M^2\Omega+M(1-\Omega)]\big),
\end{align}
where $\Omega=\common$. By assuming the total number of configurable elements of the whole BD-RIS is $Z=L\ml$, (\ref{tobeexpressed}) can be further expressed as
\begin{align}
\mathcal{R}=\frac{1}{K}\sum_{l=1}^{L} \sum_{k=1}^{K_{l}}\log_2 \qty(1 + A\qty[\frac{ \frac{Z^2}{L^2}\Omega+\frac{Z}{L}(1-\Omega)}{\qty(1-\cos\frac{\pi}{L})^2}]),
\end{align}
where $A=\frac{\rho_{\U}\lambda^{4}G_{t}G_{r}}{4^{3}\pi^{4}d_a^{\eta_{a}}d_b^{\eta_b}}$. 
Note that, $\cos\frac{\pi}{L}$ $\in [0, 1]$, for $L \in \qty{2,3,...,6}$. 
For instance, when the total number of configurable elements $Z$ are distributed into $L=2$ and $L=6$ sectors, we have the expressions in (\ref{se2}) and (\ref{se6}), respectively. With that, the proof is complete. 

\linespread{1.1}
\vspace{0mm}
\sloppy
\bibliographystyle{IEEEtran}
\bibliography{IEEEabrv,References}
\pagestyle{empty}

\end{document}